\documentclass[nofootinbib,preprint,superscriptaddress,prC]{revtex4-2}

\usepackage[utf8]{inputenc}
\usepackage{amsmath}
\usepackage{natbib}
\usepackage{xspace}
\usepackage{cancel}
\usepackage{hyperref}
\usepackage{graphicx}

\newcommand{\Ap}{\bar{\mathcal{A}}^+}
\newcommand{\Am}{\bar{\mathcal{A}}^-}
\newcommand{\Gc}{\bar{\mathcal{G}}}
\newcommand{\Gct}{\tilde{\bar{\mathcal{G}}}}
\newcommand{\Nh}{\hat{N}}
\newcommand{\pp}{\mathbf{p}_+}
\newcommand{\pn}{\mathbf{p}_-}
\newcommand{\sv}{\vec{\sigma}}
\newcommand{\tv}{\vec{\tau}}
\newcommand{\Cone}{\bar{\mathcal{C}}^{({}^3\!S_1-{}^1\!P_1)}}
\newcommand{\Ctwo}{\bar{\mathcal{C}}^{({}^1\!S_0-{}^3\!P_0)}_{(\Delta I=0)}}
\newcommand{\Cthree}{\bar{\mathcal{C}}^{({}^1\!S_0-{}^3\!P_0)}_{(\Delta I=1)}}
\newcommand{\Cfour}{\bar{\mathcal{C}}^{({}^1\!S_0-{}^3\!P_0)}_{(\Delta I=2)}}
\newcommand{\Cfive}{\bar{\mathcal{C}}^{({}^3\!S_1-{}^3\!P_1)}}
\newcommand{\vect}[1]{\vec{\mathbf{#1}}}
\newcommand{\nablad}{\stackrel{\leftrightarrow}{\nabla}}
\newcommand{\vectS}[1]{\vec{\boldsymbol{#1}}}
\newcommand{\gTSOTPO}{\bar{g}^{{}^{3}\!S_{1}-{}^{3}\!P_{1}}}
\newcommand{\gTSOOPO}{\bar{g}^{{}^{3}\!S_{1}-{}^{1}\!P_{1}}}
\newcommand{\gDIZ}{\bar{g}^{{}^{1}\!S_{0}-{}^{3}\!P_{0}}_{(\Delta I=0)}}
\newcommand{\gDIO}{\bar{g}^{{}^{1}\!S_{0}-{}^{3}\!P_{0}}_{(\Delta I=1)}}
\newcommand{\gDIT}{\bar{g}^{{}^{1}\!S_{0}-{}^{3}\!P_{0}}_{(\Delta I=2)}}
\newcommand{\Tv}{\cancel{T}}
\newcommand{\TPV}{\cancel{T}\cancel{P}}
\newcommand{\cpv}{\cancel{CP}}
\newcommand{\EFT}{$\mathrm{EFT}(/\!\!\!\pi)$\xspace}
\newcommand{\DS}{^2\!S_{\frac{1}{2}}}
\newcommand{\NN}{N\!N}
\newcommand\CG[6]{C_{#1,#2;#3}^{#4,#5;#6}}

\begin{document}

\title{Time-Reversal-Invariance Violation in the $N\!d$ System and Large-$N_C$}
\author{Anna C. Bowman}
\email{ABowman@Knights.ucf.edu}
\affiliation{Department of Physics, Stetson University, DeLand, FL 32723}
\affiliation{Department of Physics, University of Central Florida, Orlando, FL 32826}

\author{Jared Vanasse}
\email{jvanass3@fitchburgstate.edu}
\affiliation{Fitchburg State University,
Fitchburg, MA 160 Pearl St. 01420}
\affiliation{Department of Physics, Stetson University, DeLand, FL 32723}

\date{\today}

\begin{abstract}
    A minimal set of five low energy constants (LECs) for time-reversal and parity violating ($\TPV$) nucleon-nucleon ($\NN$) interactions at low energies ($E\!<\!m_{\pi}^2/M_N$) is given.  Using a large-$N_C$ (number of colors in QCD) analysis we show that one linear combination of LECs is $\mathcal{O}(N_C)$, three LECs are $\mathcal{O}(N_C^{0})$, and one linear combination of LECs is $\mathcal{O}(N_C^{-1})$.  We also calculate the $\TPV$ observables of neutron spin rotation through a polarized deuteron target and a spin correlation coefficient in nucleon-deuteron scattering using pionless effective field theory.  Using the large-$N_C$ analysis we show that the spin correlation coefficient and the neutron spin rotation are predominantly determined by the same two LECs in the large-$N_C$ basis.
\end{abstract}

\maketitle
\newpage

\section{Introduction}

Time-reversal ($T$) symmetry is an invariance of the laws of physics under the transformation $t\to-t$.  In the Standard Model (SM) the only known source of $T$ violation ($\Tv$) that manifests in nucleon-nucleon ($\NN$) interactions comes from a complex phase in the CKM matrix~\cite{Kobayashi:1973fv}.  In QCD the $\bar{\theta}$ term~\cite{tHooft:1976rip} also gives rise to $T$ violating $\NN$ interactions but is currently consistent with zero.  The unnatural smallness of the $\bar{\theta}$ term is known as the ``strong CP problem".  One possible solution is provided by the Peccei-Quinn mechanism~\cite{Peccei:1977hh}, which leads to the creation of axions a possible dark matter candidate~\cite{Weinberg:1977ma,Wilczek:1977pj}.  By the CPT theorem $\Tv$ is equivalent to  the violation of the product of charge-symmetry ($C$) (symmetric under interchange of particle and anti-particle) and parity ($P$) (symmetric under change in sign of coordinates).  $CP$-violation ($\cpv$) is a necessary condition to obtain a matter antimatter asymmetry in the universe~\cite{Sakharov:1967dj}.  However, the amount of $\cpv$ in the SM is not enough to account for the observed matter antimatter asymmetry in the universe~\cite{Canetti:2012zc}.  Thus, it is expected beyond the SM (BSM) physics must have further sources of $\cpv$.  

BSM physics can be encoded in an effective field theory (EFT) that respects SM symmetries known as SM EFT.  BSM theories should reduce to the SM at low energies and can be matched to the SM EFT by integrating out heavy degrees of freedom.  Different BSM theories will give different values for the low energy constants (LECs) of higher dimension ($d>4$) non renormalizable operators in SM EFT.  The $d=6$ $\cpv$ SM EFT operators have been delineated in Ref.~\cite{DeRujula:1990db}.  By using  renormalization group (RG) and EFT techniques the $d=6$ $\cpv$ operators can be run down to $\Lambda_{\mathrm{QCD}}\sim 1$ GeV and matched to relevant QCD operators.  This has been done at tree level~\cite{deVries:2012ab}.  Below $\Lambda_{\mathrm{QCD}}$ the matching of QCD operators to chiral EFT ($\chi$EFT) is nontrivial due to the nonperturbative nature of QCD, but can in principle be done through lattice QCD~\cite{Shintani:2015vsx}.  Despite not being able to directly match the $d=6$ SM EFT $\cpv$ operators to $\cpv$ operators in $\chi$EFT they have been related to each other by using the pattern in which the operators break chiral symmetry~\cite{Mereghetti:2010tp,deVries:2012ab}.  

At low energies ($E\lesssim m_{\pi}^2/M_N$) interactions between nuclei can be described in a series of contact interactions between nuclei known as pionless EFT (\EFT).  \EFT has been used to great success to describe the static properties of few nucleon systems and interactions between light nuclei at low energies (See Refs.~\cite{Beane:2000fx} and~\cite{Vanasse:2016jtc} for reviews).  In \EFT $T$ and $P$ violating ($\TPV$) $\NN$ interactions are described by five independent LECs~\cite{Maekawa:2011vs}, which can be matched to $\chi$EFT providing a connection to BSM physics where the only weak link in the chain is the matching of $\chi$EFT to QCD.  These five LECs must be determined from experiment or fundamental interactions through lattice QCD.  Experiments involving heavy nuclei offer the possibility of an enhanced $T$-violating signal due to closely spaced nuclear levels that behave oppositely under $T$-symmetry and seem ideal candidates to determine the five LECs~\cite{Gudkov:1991qg}.  However, calculating properties of heavy nuclei is difficult and to cleanly extract the LECs from experiment it is preferable to do experiments on few-nucleon systems as is being carried out for $P$-violating (PV) $\NN$ interactions~\cite{Snow:2016zyq}.  Given that there are five LECs it would be desirable to further distinguish the relative size of these LECs.  Such a scheme is provided by a large-$N_C$ analysis in QCD~\cite{tHooft:1973alw,Witten:1979kh} in which the number of colors ($N_C$) in QCD is used as an expansion parameter.  This analysis has been carried out for all $T$-violating $\NN$ operators to order $N_C^{-1}$~\cite{Samart:2016ufg}.  Below we show how this general analysis reduces to five LECs and find the large-$N_C$ scaling of these LECs in analogy to what has been done in the PV sector~\cite{Schindler:2015nga}.

Electric dipole moments (EDMs) of nuclei and neutral atoms violate both $T$ and $P$ and are currently of great interest in searches for $\Tv$.  The neutron (proton) EDM has been measured to $|d_n|<2.9\times 10^{-13}e$ fm~\cite{Baker:2006ts} ($|d_p|<7.9\times 10^{-12}e$ fm), while the SM prediction is estimated at $|d_n|\sim|d_p|\sim10^{-19} e$ fm~\cite{Seng:2014lea}.  The current best bounds for the proton EDM come from the EDM bounds on $^{199}$Hg~\cite{Griffith:2009zz}.  Future experiments expect to bring neutron EDM measurements down two orders of magnitude~\cite{Ban:2006qp,Kumar:2013qya,Serebrov:2018lxv}.  Proposed charge storage ring experiments could in principle measure the proton, deuteron, and $^3$He EDMs to a precision of $\sim\!\!10^{-16} e$ fm~\cite{Semertzidis:2011qv,Pretz:2013us}.  $\chi$EFT has been used to calculate the light $A<4$ nuclear EDMs~\cite{deVries:2010ah,deVries:2011an,deVries:2012ab,Bsaisou:2012rg,Bsaisou:2014zwa,Gnech:2019dod}, as well as phenomenological and hybrid models \cite{Yamanaka:2015ncb,Song:2012yh}, and \EFT~\cite{Yang:2020ges}.  Measurements from several light nuclear EDMs would allow for the disentanglement of contributions  from different $d=4$ and $d=6$ SM EFT $\cpv$ operators and make clearer the picture of BSM physics in the $\cpv$ sector.

Another avenue to find $\Tv$ in nuclei complementary to EDM searches is through neutron spin rotation experiments on polarized nuclear targets and spin-correlation experiments with nucleon-nucleus scattering.  These observables have been previously investigated in the neutron-deuteron ($nd$) system with a \EFT $\TPV$ $\NN$ potential~\cite{Song:2011sw}.  However, these calculations used the so called hybrid method in which strong interactions were given by the phenomenological potentials of AV18+UIX~\cite{Wiringa:1994wb,Pudliner:1995wk}, while $\TPV$ $\NN$ interactions were given by \EFT.  In this work we calculate these observables in a completely consistent \EFT framework in which \EFT is used both for the strong and $\TPV$ $\NN$ interactions.  A consistent \EFT calculation allows for the full machinery of error estimation in EFT to be properly utilized.  We only calculate to leading-order (LO) in \EFT since a next-to-leading order (NLO) calculation will likely require the inclusion of a $\TPV$ three-body force as this is the case for the analogous PV $\NN$ interactions~\cite{Vanasse:2018buq}.  In addition we analyze the constraints on these observables placed by large-$N_C$.

This paper is organized as follows.  In Sec.~\ref{sec:lag} the LO strong and $\TPV$ \EFT Lagrangian in the two and three-nucleon sector is given.  Section~\ref{sec:largenc} derives the large-$N_C$ counting of the five $\TPV$ LECs in \EFT.  The calculation of the nucleon-deuteron ($N\!d$) scattering amplitude including $\TPV$ interactions is discussed in Sec.~\ref{sec:threebody}.  Section~\ref{sec:observables} gives the $\TPV$ observables in the $N\!d$ system in terms of partial wave amplitudes and Sec. \ref{sec:results} discusses the results of the calculated observables.  Finally, we conclude in Sec. \ref{sec:conclusion}.

\section{\label{sec:lag}Lagrangian}

The LO Lagrangian in \EFT is given by
\begin{align}
&\mathcal{L}=\hat{N}^{\dagger}\left(i\partial_{0}+\frac{\vect{\nabla}^{2}}{2M_{N}}\right)\hat{N}+\hat{t}_{i}^{\dagger}\Delta_{t}\hat{t}_{i}+\hat{s}_{a}^{\dagger}\Delta_{s}\hat{s}_{a}\\\nonumber
&-y\left[\hat{t}_{i}^{\dagger}\hat{N} ^{T}P_{i}\hat{N}+\hat{s}_{a}^{\dagger}\hat{N}^{T}\bar{P}_{a}\hat{N}+\mathrm{H.c.}\right]\\\nonumber
&+\frac{y^{2}M_{N}H_{\mathrm{LO}}(\Lambda)}{3\Lambda^{2}}\left[\hat{t}_{i}(\sigma_{i}\hat{N})-\hat{s}_{a}(\tau_{a}\hat{N})\right]^{\dagger}\left[\hat{t}_{i}(\sigma_{i}\hat{N})-\hat{s}_{a}(\tau_{a}\hat{N})\right],
\end{align}
where $\hat{N}$, $\hat{t}_i$, and $\hat{s}_a$ are the nucleon, spin-triplet (deuteron), and spin-singlet dibaryon field respectively.   $P_i=\frac{1}{\sqrt{8}}\sigma_2\sigma_i\tau_2$ ($\bar{P}_a=\frac{1}{\sqrt{8}}\sigma_2\tau_2\tau_a$) projects out the spin-triplet iso-singlet (spin-singlet iso-triplet) combination of nuclei.  The two-body parameters are fit to the deuteron binding momentum $\gamma_t=45.7025$~MeV and the $^1\!S_0$ virtual bound state momentum $\gamma_s=-7.890$~MeV yielding~\cite{Griesshammer:2004pe}
\begin{equation}
    \Delta_t=\gamma_t-\mu\quad,\quad\Delta_s=\gamma_s-\mu\quad,\quad y^2=\frac{4\pi}{M_N}.
\end{equation}
$\mu$ is a scale that comes from using dimensional regularization with power divergence subtraction~\cite{Kaplan:1996xu,Kaplan:1998tg} and physical observables are independent of $\mu$.  $H_{\mathrm{LO}}(\Lambda)$, the LO three-body force~\cite{Bedaque:1998kg}, is fit to the $\DS$ $nd$ scattering length $a_{nd}=0.65$~fm~\cite{Dilg:1971gqb}.  The scale $\Lambda$ comes from regulating momentum integrals with a hard cutoff.  For details of fitting the three-body force see Ref.~\cite{Vanasse:2015fph}.  The LO $\NN$ scattering amplitude is given by an infinite sum of diagrams and is related to the dibaryon propagator~\cite{Griesshammer:2004pe}
\begin{equation}
    D_{\{t,s\}}(E,p)=\frac{1}{\sqrt{\frac{3}{4}p^2-M_NE}-\gamma_{\{t,s\}}},
\end{equation}
where $t$ ($s$) is the spin-triplet (spin-singlet) dibaryon propagator.  Taking the residue of the spin-triplet dibaryon propagator about the bound state pole gives the LO deuteron wavefunction renormalization
\begin{equation}
    Z_{\mathrm{LO}}=\frac{2\gamma_t}{M_N}.
\end{equation}
\indent The LO Lagrangian for two-body $\TPV$ violating interactions in \EFT is given by
\begin{align}
\label{eq:PTVLagdib}
\mathcal{L}_{\TPV}=-&\left[\gTSOOPO\hat{t}_{i}^{\dagger}\left(\Nh^{T}\sigma_{2}\tau_2\nablad_{i}\!\Nh\right)\right.\\\nonumber
&+\gDIZ \hat{s}_{a}^{\dagger}\left(\Nh^{T}\sigma_{2}\vectS{\sigma}\cdot\tau_{2}\tau_{a}\nablad\!\Nh\right)\\\nonumber
&+\gDIO\epsilon^{3ab}\hat{s}_{a}^{\dagger}\left(\Nh^{T}\sigma_{2}\vectS{\sigma}\cdot\tau_{2}\tau_{b}i\nablad\!\Nh\right)\\\nonumber
&+\gDIT\mathcal{I}^{ab}\hat{s}_{a}^{\dagger}\left(\Nh^{T}\sigma_{2}\vectS{\sigma}\cdot\tau_{2}\tau_{b}\nablad\!\Nh\right)\\\nonumber
&\left.+\gTSOTPO\epsilon^{ijk}\hat{t}_{i}^{\dagger}\left(\Nh^{T}\sigma_{2}\sigma^{k}\tau_{2}\tau_{3}i\nablad{}^{\!j}\Nh\right)\right]+\mathrm{H.c.},
\end{align}
where $\mathcal{I}^{ab}=\mathrm{diag}(1,1,-2)$ projects out an isotensor.  This is analogous to the Lagrangian for PV interactions but with additional factors of $i$ or $-i$~\cite{Phillips:2008hn}.  The operator $\nablad$ is defined via $b\nablad a=b(\nabla a)-(\nabla b)a$.  To distinguish $\TPV$ LECs from similar PV LECs we place a bar over them.

\section{\label{sec:largenc}Large-$N_C$}

The values of the five $\TPV$ LECs are entirely unconstrained by experiment.  However, the large-$N_C$ expansion of QCD allows for the discernment of the relative size of these LECs. The most general $\TPV$ $\NN$ potential with a single power of momentum is given by
\begin{align}
    \label{eq:Vnonmin}
    V^{\mathrm{nonmin}}=&\Am_1\pn\cdot i(\sv_1-\sv_2)\\\nonumber
    &+\Ap_1\pp\cdot (\sv_1\times\sv_2)\\\nonumber
    &+\Am_2\pn\cdot i(\sv_1\tau_1^3-\sv_2\tau_2^3)\\\nonumber
    &+\frac{1}{2}\Ap_2\pp\cdot (\sv_1\times\sv_2)(\tau_1+\tau_2)^3\\\nonumber
    &+\Am_3\pn\cdot i(\sv_1-\sv_2)\tv_1\cdot\tv_2\\\nonumber
    &+\Ap_3\pp\cdot (\sv_1\times\sv_2)\tv_1\cdot\tv_2\\\nonumber
    &+\Am_4\pn\cdot i(\sv_1\tau_2^3-\sv_2\tau_1^3)\\\nonumber
    &+\Am_5\pn\cdot i(\sv_1-\sv_2)\mathcal{I}_{ab}\tau_1^a\tau_2^b\\\nonumber
    &+\Ap_5\pp\cdot (\sv_1\times\sv_2)\mathcal{I}_{ab}\tau_1^a\tau_2^b\\\nonumber
    &-\frac{1}{2}\Ap_6\pp\cdot i(\sv_1+\sv_2)i(\tau_1\times\tau_2)^3,
\end{align}
where
\begin{equation}
    \mathbf{p}_{\pm}=\mathbf{p}'\pm\mathbf{p},
\end{equation}
and
\begin{equation}
    \mathbf{p}'=\mathbf{p}_1'-\mathbf{p}_2'\quad,\quad\mathbf{p}=\mathbf{p}_1-\mathbf{p}_2.
\end{equation}
$\mathbf{p}_1$ ($\mathbf{p}_1'$) and $\mathbf{p}_2$ ($\mathbf{p}_2'$) are the momenta of the incoming (outgoing) nucleons.  The large-$N_C$ scaling of the coefficients derived in Ref.~\cite{Samart:2016ufg} is
\begin{align}
    &\Ap_1\sim N_C^{-2}, && \Am_1\sim N_C^{0}  \\\nonumber
    &\Ap_2\sim N_C^{-1}, && \Am_2\sim N_C  \\\nonumber
    &\Ap_3\sim N_C^{0}, && \Am_3\sim N_C^{0} \\\nonumber
    & && \Am_4\sim N_C^{-1} \\\nonumber
    &\Ap_5\sim N_C^{0}, && \Am_5\sim N_C^{0} \\\nonumber
    &\Ap_6\sim N_C^{-1}, &&  \\\nonumber
\end{align}
Many of the operators in Eq.~(\ref{eq:Vnonmin}) are interrelated via Fierz transformations and can be simplified to a set of five independent operators.  This reduction has been carried out previously by Girlanda for PV operators and obtained a Lagrangian with five LECs~\cite{Girlanda:2008ts}.  The $\TPV$ Lagrangian can be obtained from the Girlanda Lagrangian for PV operators by simply interchanging $\pp$ and $\pn$ and adding factors of $i$ or $-i$ giving
\begin{align}
\label{eq:GirLag}
\mathcal{L}_{\TPV}^{min}=&\Gc_1(\Nh^{\dagger}\vec{\sigma}\Nh\cdot \nabla(\Nh^{\dagger}\Nh)-\Nh^{\dagger}\Nh\nabla\cdot(\Nh^{\dagger}\vec{\sigma}\Nh))\\\nonumber
&-\Gct_1\epsilon_{ijk}\Nh^{\dagger}\sigma^{i}\Nh\Nh^{\dagger}\sigma^{k}i\nablad{}^{\!j}\Nh\\\nonumber&
-\Gc_2\epsilon_{ijk}[\Nh^{\dagger}\tau^{3}\sigma^{i}\Nh\Nh^{\dagger}\sigma^{k}i\nablad{}^{\!j}\Nh+\Nh^{\dagger}\sigma^{i}\Nh\Nh^{\dagger}\tau^{3}\sigma^{k}i\nablad{}^{\!j}\Nh]\\\nonumber
&-\Gc_5\mathcal{I}_{ab}\epsilon_{ijk}\Nh^{\dagger}\tau^{a}\sigma^{i}\Nh\Nh^{\dagger}\tau^{b}\sigma^{k}i\nablad{}^{\!j}\Nh\\\nonumber
&+\Gc_6\epsilon_{ab3}\Nh^{\dagger}\tau^{a}i\nablad\! \Nh\cdot(\Nh^{\dagger}\tau^{b}\vec{\sigma}\Nh).
\end{align}
The resulting potential from this set of operators is
\begin{align}
    \label{eq:Vmin}
    V^{\mathrm{min}}=&-\Gc_1\pn\cdot i(\sv_1-\sv_2)\\\nonumber
    &-\Gct_1\pp\cdot (\sv_1\times\sv_2)\\\nonumber
    &-\Gc_2\pp\cdot (\sv_1\times\sv_2)(\tau_1+\tau_2)^3\\\nonumber
    &-\Gc_5\pp\cdot (\sv_1\times\sv_2)\mathcal{I}_{ab}\tau_1^a\tau_2^b\\\nonumber
    &+\frac{1}{2}\Gc_6\pp\cdot(\sv_1+\sv_2)(\tau_1\times\tau_2)^3.
\end{align}
Using Fierz rearrangements the coefficients of the over complete potential Eq.~(\ref{eq:Vnonmin}) can be related to the coefficients of the minimal Girlanda potential yielding
\begin{align}
    & \Gc_1=-\Am_1+\Am_3+2\Ap_3\\\nonumber
    & \Gct_1=-\Ap_1+2\Am_3+\Ap_3\\\nonumber
    & \Gc_2=-\frac{1}{2}\left(\Ap_2-\Am_2-\Am_4\right)\\\nonumber
    & \Gc_5=-\left(\Ap_5-\Am_5\right)\\\nonumber
    & \Gc_6=-\Ap_6-\Am_2+\Am_4,
\end{align}
which gives the large-$N_C$ scaling 
\begin{align}
    &\Gc_2\sim\Gc_6\sim N_C\\\nonumber
    &\Gc_1\sim\Gct_1\sim\Gc_5\sim N_c^0
\end{align}
and the relation
\begin{equation}
    \label{eq:relation}
    \Gc_2=-\frac{1}{2}\Gc_6,
\end{equation}
which holds to order $\mathcal{O}(N_C^{-1})$.  An alternative basis for the five independent LECs is the partial wave basis in which the incoming and outgoing partial waves of the nucleons are manifest.  The Lagrangian in the partial wave basis is
\begin{align}
\mathcal{L}_{\TPV}=  -  & \left[ \Cone \left(\Nh^T\sigma^2  \vec{\sigma} \tau^2 \Nh \right)^\dagger
\cdot  \left(\Nh^T \sigma^2  \tau^2 \nablad\! \Nh\right) \right. \\\nonumber
& +\Ctwo \left(\Nh^T\sigma^2 \tau^2 \vec{\tau} \Nh\right)^\dagger
\left(\Nh^T\sigma^2  \vec{\sigma} \cdot \tau^2 \vec{\tau} \nablad\!  \Nh\right) \\\nonumber
& +\Cthree \ \epsilon_{3ab} \left(\Nh^T\sigma^2 \tau^2 \tau^a \Nh\right)^\dagger
\left(\Nh^T \sigma^2   \vec{\sigma} \cdot \tau^2 \tau^b i\nablad\! \Nh\right) \\\nonumber
& +\Cfour \ \mathcal{I}_{ab} \left(\Nh^T\sigma^2 \tau^2 \tau^a \Nh\right)^\dagger
\left(\Nh^T \sigma^2  \vec{\sigma} \cdot \tau^2 \tau^b  \nablad\! \Nh\right) \\\nonumber
& +\left. \Cfive \ \epsilon_{ijk} \left(\Nh^T\sigma^2 \sigma^i \tau^2 \Nh\right)^\dagger
\left(\Nh^T \sigma^2 \sigma^k \tau^2 \tau^3 i\nablad{}^{\!j} \Nh\right) \right] +\mathrm{H.c.} \, ,
\end{align}
which is nearly identical to the Lagrangian for the PV $\NN$ potential~\cite{Phillips:2008hn} except for additional factors of $i$ or $-i$.  LECs in the partial wave basis can be related to LECs in the Girlanda basis using Fierz rearrangements~\cite{vanasse2012parity} or techniques described in Ref.~\cite{Phillips:2008hn} yielding
\begin{align}
    &\Cone=-\frac{1}{4}(\Gc_1+\Gct_1)\\\nonumber
    &\Ctwo=-\frac{1}{4}(\Gc_1-\Gct_1)\\\nonumber
    &\Cthree=-\frac{1}{2}\Gc_2\\\nonumber
    &\Cfour=-\frac{1}{2}\Gc_5\\\nonumber
    &\Cfive=-\frac{1}{4}\Gc_6.
\end{align}
From the matching and the large-$N_C$ scaling of the Girlanda LECs the large-$N_C$ scaling of the LECs in the partial wave basis is
\begin{align}
    &\Cfive\sim\Cthree\sim N_C\\\nonumber
    &\Cone\sim\Ctwo\sim\Cfour\sim N_c^0,
\end{align}
where
\begin{align}
    \label{eq:relationpartial}
    \Cthree=-\Cfive,
\end{align}
to order $\mathcal{O}(N_C^{-1})$.  Finally, we match the partial wave basis LECs to the dibaryon formalism LECs in Eq.~(\ref{eq:PTVLagdib}).  This can be done by either a simple matching calculation or a Gaussian integration over the dibaryon fields.  The matching yields the relation~\cite{Schindler:2009wd}
\begin{equation}
    \label{eq:matching}
    \frac{\bar{g}^{(X-Y)}}{y}=\sqrt{8}\frac{\bar{\mathcal{C}}^{(X-Y)}}{\mathcal{C}^{(X)}_0},
\end{equation}
where $X$ ($Y$) is $^{1}\!S_0$ or $^3\!S_1$ ($^1\!P_0$, $^3\!P_0$, or $^3\!P_1$) and subscripts of $\Delta I=0,1,2$ not shown are understood to be the same on both sides.  Partial wave basis LECs $C^{(X)}_0$ are given by the Lagrangian
\begin{equation}
    \mathcal{L}_2=-\mathcal{C}_0^{(^1\!S_0)}(\Nh^T\bar{P}_a\hat{N})^{\dagger}\Nh^T\bar{P}_a\Nh-\mathcal{C}_0^{(^3\!S_1)}(\Nh^T P_i\Nh)^{\dagger}\Nh^T P_i\Nh.
\end{equation}
Large-$N_C$ shows that at $\mathcal{O}(N_C)$
\begin{equation}
    \mathcal{C}_0^{(^3\!S_1)}=\mathcal{C}_0^{(^1\!S_0)}
\end{equation}
and this holds to $\mathcal{O}(N_C^{-1
})$~\cite{Kaplan:1995yg}.  Following Ref.~\cite{Vanasse:2011nd} we define the coefficients
\begin{equation}
    \bar{g}_1=\frac{\gTSOOPO}{y},\bar{g}_2=\frac{\gTSOTPO}{y},\bar{g}_3=\frac{\gDIZ}{y},\bar{g}_4=\frac{\gDIO}{y},\bar{g}_5=\frac{\gDIT}{y}.
\end{equation}
Using Eq.~(\ref{eq:matching}) and the large $N_C$ scaling of the LECs in the partial wave basis we define the large-$N_C$ basis of dibaryon LECs as
\begin{align}
    &\bar{g}_1^{(N_C)}=\frac{1}{2}(\bar{g}_2-\bar{g}_4) && \mathrm{LO}(\mathcal{O}(N_C))\\\nonumber
    &\bar{g}_2^{(N_C^0)}=\bar{g}_1,\bar{g}_3^{(N_C^0)}=\bar{g}_3,\bar{g}_4^{(N_C^0)}=\bar{g}_5 && \mathrm{NLO}(\mathcal{O}(N_C^0))\\\nonumber
    &\bar{g}_5^{(N_C^{-1})}=\frac{1}{2}(\bar{g}_2+\bar{g}_4) && \mathrm{NNLO}(\mathcal{O}(N_C^{-1})),
\end{align}
where $\bar{g}_5^{(N_C^{-1})}$ is the next-to-next-to leading-order (NNLO) $\mathcal{O}(N_C^{-1})$  in large-$N_C$ LEC. The $\mathcal{O}(N_C)$ and $\mathcal{O}(N_C^{-1})$ combination of LECs come from the use of Eq.~(\ref{eq:relationpartial}).

\section{\label{sec:threebody}Three-Body System}

\begin{figure}[hbt!]
    \centering
    \includegraphics[width=100mm]{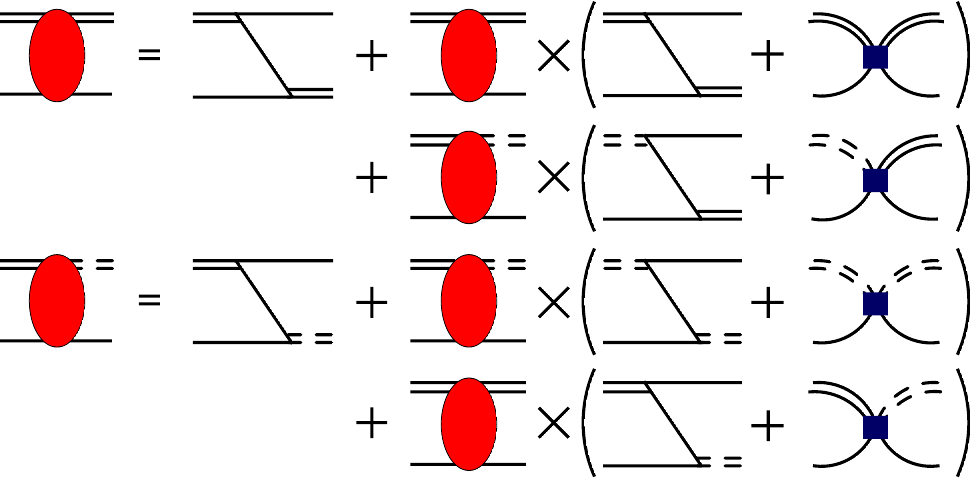}
    \caption{Diagrammatic representation of integral equations for the parity and time-reversal conserving LO $N\!d$ scattering amplitude. The single line is a nucleon, solid double line a spin-triplet dibaryon, dashed double line a spin-singlet dibaryon, solid square the LO three-body force, and the red oval with solid double lines is the LO $N\!d$ scattering amplitude.}
    \label{fig:PCIntegralEq}
\end{figure}

The LO $N\!d$ scattering amplitude is given by an infinite sum of diagrams in \EFT.  This sum of diagrams is solved via the integral equation represented diagrammatically in Fig.~\ref{fig:PCIntegralEq}.  
Projecting this integral equation in the total angular momentum basis yields the LO time-reversal and parity conserving ($T\!P$) integral equation
\begin{align}
\label{eq:PCInegraleq}
{\mathbf{t}}^{J}_{L'S',LS}(k,p,E)=&{\mathbf{K}}^{J}_{L'S',LS}(k,p,E)\mathbf{v}_{p}\\\nonumber
&+\sum_{L'',S''}{\mathbf{K}}^{J}_{L'S',L''S''}(q,p,E)\mathbf{D}\left(E,q\right)\otimes_q{\mathbf{t}}^{J}_{L''S'',LS}(q,p,E),
\end{align}
where $L$ ($L'$) is the incoming (outgoing) orbital angular momentum between nucleon and deuteron, $S$ ($S'$) is the total incoming (outgoing) spin angular momentum in the $N\!d$ system, and $J$ is the total angular momentum.  $k$ ($p$) is the magnitude of the incoming (outgoing) on-shell (off-shell) momentum of the nucleon in the center-of-mass (c.m.) frame, where the on-shell condition is $E=\frac{3k^2}{4M_N}-\frac{\gamma_t^2}{M_N}$, with $E$ being the total energy of the $N\!d$ system. The kernel ${\mathbf{K}}^{J}_{L'S',LS}(k,p,E)$ is a matrix in cluster-configuration (c.c.) space~\cite{Griesshammer:2004pe} defined by~\cite{Vanasse:2018buq}
\begin{align}
\label{eq:PCKernel}
&{\mathbf{K}}^{J}_{L'S',LS}(k,p,E)=\\\nonumber
&\delta_{LL'}\delta_{SS'}(-1)^{L}\left\{
\begin{array}{cc}
\frac{2\pi}{kp}Q_{L}\left(\frac{k^{2}+p^{2}-M_{N}E-i\epsilon}{kp}\right)\left(\!\!\begin{array}{rr}
1 & -3\\[-1mm]
-3 & 1
\end{array}\right)+\frac{4\pi H_{\mathrm{LO}}(\Lambda)}{\Lambda^{2}}\delta_{L0}
\left(\!\!\begin{array}{rr}
1 & -1\\[-1mm]
-1 & 1
\end{array}\right) & ,S=\frac{1}{2}\\
-\frac{4\pi}{kp}Q_{L}\left(\frac{k^{2}+p^{2}-M_{N}E-i\epsilon}{kp}\right)\left(\begin{array}{rr}
1 & 0\\[-1mm]
0 & 0 
\end{array}\right) & ,S=\frac{3}{2}
\end{array}
\right.,
\end{align}
where $Q_L(a)$ is a Legendre function of the second kind defined by
\begin{equation}
    Q_L(a)=\frac{1}{2}\int_{-1}^1\frac{P_L(x)}{x-a},
\end{equation}
and $P_L(x)$ are the standard Legendre polynomials.  $\mathbf{D}(E,q)$ is a matrix in c.c.~space given by
\begin{equation}
    \mathbf{D}(E,q)=\left(\!\!\begin{array}{cc}
    D_t(E,q) & 0\\
    0 & D_s(E,q)
    \end{array}\!\!\right),
\end{equation}
and $\mathbf{t}^{J}_{L'S',LS}(k,p,E)$ is a vector in c.c.~space defined by
\begin{equation}
    \mathbf{t}^{J}_{L'S',LS}(k,p,E)=\left(\!\!\begin{array}{c}
    {t}^{J;Nt\to Nt}_{L'S',LS}(k,p,E) \\
    {t}^{J;Nt\to Ns}_{L'S',LS}(k,p,E)
    \end{array}\!\!\right),
\end{equation}
where ${t}^{J;Nt\to Nt}_{L'S',LS}(k,p,E)$ is the $N\!d$ scattering amplitude and ${t}^{J;Nt\to Ns}_{L'S',LS}(k,p,E)$ is the unphysical scattering amplitude for a nucleon and deuteron going to a nucleon and spin singlet dibaryon.  The $\otimes_q$ notation is defined by
\begin{equation}
    A(q)\otimes_qB(q)=\frac{1}{2\pi^2}\int_0^\Lambda dqq^2A(q)B(q).
\end{equation}
Finally, $\mathbf{v}_p$ is a vector in c.c.~space that picks out spin-triplet dibaryons for the outgoing dibaryon legs and is given by
\begin{equation}
    \mathbf{v}_p=\Big(\!\begin{array}{c}
    1 \\[-3mm]
    0
    \end{array}\!\Big).
\end{equation}

\begin{figure}[hbt!]
    \centering
    \includegraphics[width=100mm]{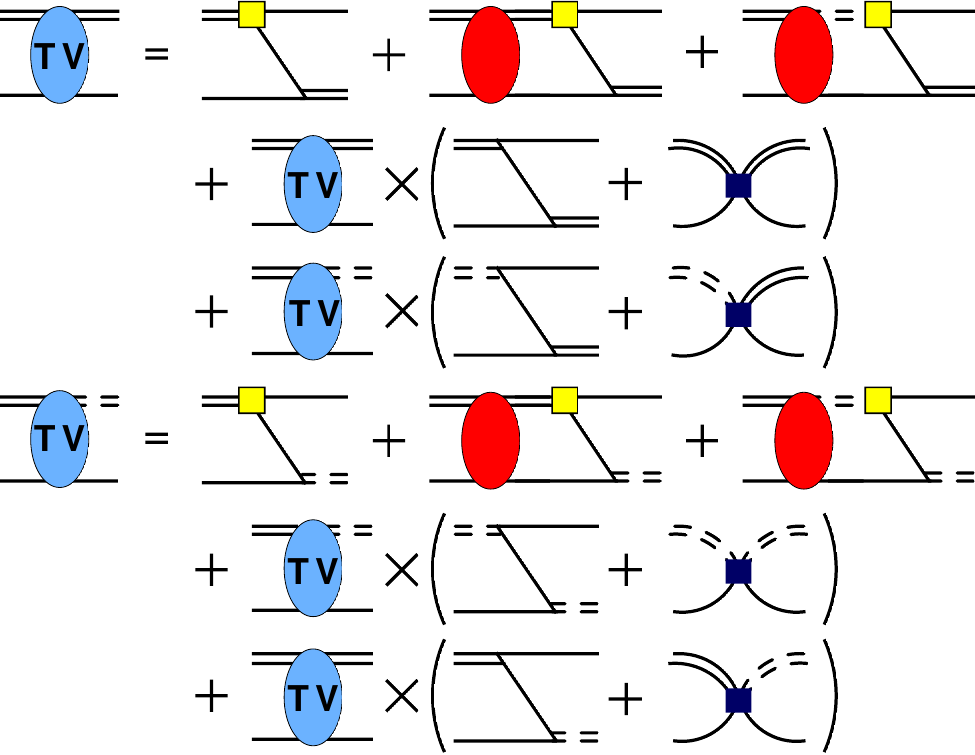}
    \caption{Diagrammatic representation of integral equations for the $\TPV$ LO $N\!d$ scattering amplitude. The light yellow square is an insertion of a $\TPV$ $\NN$ LEC, the oval with TV in it and solid double lines is the $\TPV$ LO $N\!d$ scattering amplitude, and everything else is the same as in Fig.~\ref{fig:PCIntegralEq}.}
    \label{fig:TVIntegralEq}
\end{figure}

\begin{figure}[hbt!]
    \centering
    \includegraphics[width=50mm]{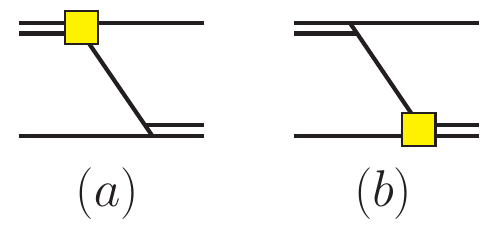}
    \caption{Diagrammatic representation of trree level contributions to the $\TPV$ LO $N\!d$ scattering amplitude.  The yellow square is an insertion of a $\TPV$ $NN$ interaction.}
    \label{fig:tree-level}
\end{figure}

The $\TPV$ $N\!d$ scattering amplitude is given by the integral equation in Fig.~\ref{fig:TVIntegralEq}.
Projecting out the integral equation in a total angular momentum basis yields
\begin{align}
\label{eq:PVintegralEq}
&{\mathbf{t}_{\TPV}}^{J}_{L'S',LS}(k,p,E)={\mathbf{K}_{\TPV}}^{J}_{L'S',LS}(k,p,E)\mathbf{v}_{p}\\\nonumber
&\hspace{1cm}+\sum_{L'',S''}{\mathbf{K}_{\TPV}}^{J}_{L'S',L''S''}(q,p,E)\otimes_q\mathbf{D}\left(E,q\right){\mathbf{t}}^{J}_{L''S'',LS}(q,p,E)\\\nonumber
&\hspace{1cm}+\sum_{L'',S''}{\mathbf{K}}^{J}_{L'S',L''S''}(q,p,E)\otimes_q\mathbf{D}\left(E,q\right){\mathbf{t}_{\TPV}}^{J}_{L''S'',LS}(q,p,E),
\end{align}
where ${\mathbf{t}_{\TPV}}^{J}_{L''S'',LS}(k,p,E)$ is a c.c.~space vector defined by
\begin{equation}
    {\mathbf{t}_{\TPV}}^{J}_{L'S',LS}(k,p,E)=\left(\!\!\begin{array}{c}
    {t_{\TPV}}^{J;Nt\to Nt}_{L'S',LS}(k,p,E) \\
    {t_{\TPV}}^{J;Nt\to Ns}_{L'S',LS}(k,p,E)
    \end{array}\!\!\right).
\end{equation}
${t_{\TPV}}^{J;Nt\to Nt}_{L'S',LS}(k,p,E)$ is the $\TPV$ $N\!d$ scattering amplitude and ${t_{\TPV}}^{J;Nt\to Ns}_{L'S',LS}(k,p,E)$ is an unphysical\, $\TPV$ scattering amplitude for a nucleon and deuteron going to a nucleon and spin-singlet dibaryon.  The $\TPV$ kernel is given by the sum of diagrams in Fig.~\ref{fig:tree-level} 
giving
\begin{equation}
    {\mathbf{K}_{\TPV}}^{J}_{L'S',LS}(k,p,E)={\mathbf{K}^{(a)}_{\TPV}}^{J}_{L'S',LS}(k,p,E)+{\mathbf{K}^{(b)}_{\TPV}}^{J}_{L'S',LS}(k,p,E)
\end{equation}
where ${\mathbf{K}^{(a)}_{\TPV}}^{J}_{L'S',LS}(k,p,E)$ (${\mathbf{K}^{(b)}_{\TPV}}^{J}_{L'S',LS}(k,p,E)$) is the contribution from diagram (a) (diagram (b)).  Diagram (a) and (b) are related by
\begin{equation}
    {\mathbf{K}^{(b)}_{\TPV}}^{J}_{L'S',LS}(k,p,E)=\left[{\mathbf{K}^{(a)}_{\TPV}}^{J}_{LS,L'S'}(p,k,E)\right]^{\dagger},
\end{equation}
where the superscript $\dagger$ is a conjugate transpose of the c.c.~space matrix.  Since the $\TPV$ Lagrangian essentially has an extra factor of $i$ as compared to the PV Lagrangian the complex conjugate results in a sign change as expected for a $T$-odd interaction.  The kernels for diagram (a) and (b) have been calculated previously for PV~\cite{Vanasse:2011nd,Griesshammer:2011md}.  The only difference between PV and $\TPV$ calculations is a factor $-i$ for diagram (a) and $i$ for diagram (b) as well as overall sign for the $\gTSOTPO$ and $\gDIO$ LEC terms.  When calculating the $\TPV$ kernel it is convenient to use a basis of LECs that can be used for both $nd$ and proton-deuteron ($pd$) interactions, such a notation was provided in~\cite{Griesshammer:2011md}, giving
\begin{align}
    &\bar{\mathcal{S}}_1=3\gTSOTPO-2\tau_3\gTSOTPO\\\nonumber
    &\bar{\mathcal{S}}_2=3\gTSOTPO+\tau_3\gTSOTPO\\\nonumber
    &\bar{\mathcal{T}}=3\gDIO-2\tau_3\gDIO.
\end{align}
Using this notation the $\TPV$ kernel for each partial wave channel of interest is
\begin{align}
    &\mathbf{K}{_{\TPV}}^{\frac{1}{2}}_{1\frac{1}{2}, 0 \frac{1}{2}}(k,p,E)=-\frac{4 \pi i \sqrt{2}}{3kp}\left\{pQ_0(a)\left(\!\!\begin{array}{cc}
\bar{\mathcal{S}}_1&-2\bar{\mathcal{T}}-\bar{\mathcal{S}}_1\\
\bar{\mathcal{T}}+2\bar{\mathcal{S}}_1&-\bar{\mathcal{T}}
\end{array}\!\!\right)\right.\\[2mm]\nonumber
&\left.\hspace{2cm}-kQ_1(a)\left(\!\!\begin{array}{cc}
\bar{\mathcal{S}}_1 &\bar{\mathcal{T}}+2\bar{\mathcal{S}}_1\\
-2\bar{\mathcal{T}}-\bar{\mathcal{S}}_1&-\bar{\mathcal{T}}\\
\end{array}\!\!\right)\right\},
\end{align}
\begin{align}
    &\mathbf{K}{_{\TPV}}^{\frac{1}{2}}_{1\frac{3}{2}, 0 \frac{1}{2}}(k,p,E)=\frac{8 \pi i} {3kp}\left\{pQ_0(a)\left(\!\!\begin{array}{cc}
\frac{4\bar{\mathcal{S}}_1-7\bar{\mathcal{S}}_2}{3}&4\bar{\mathcal{T}}-\bar{\mathcal{S}}_2\\
0&0
\end{array}\!\!\right)\right.\\[2mm]\nonumber
&\left.\hspace{2cm}-2kQ_1(a)\left(\!\!\begin{array}{cc}
\frac{4\bar{\mathcal{S}}_2-\bar{\mathcal{S}}_1}{3}&\bar{\mathcal{S}}_2-\bar{\mathcal{T}}\\
0&0\\
\end{array}\!\!\right)\right\},
\end{align}
\begin{align}
    &\mathbf{K}{_{\TPV}}^{\frac{3}{2}}_{1 \frac{1}{2}, 0 \frac{3}{2}}(k,p,E)=-\frac{4\sqrt{2}\pi i} {3kp}\left\{2pQ_0(a)\left(\!\!\begin{array}{cc}
\frac{4\bar{\mathcal{S}}_2-\bar{\mathcal{S}}_1}{3}&0\\
\bar{\mathcal{S}}_2-\bar{\mathcal{T}}&0
\end{array}\!\!\right)\right.\\[2mm]\nonumber
&\left.\hspace{2cm}-kQ_1(a)\left(\!\!\begin{array}{cc}
\frac{4\bar{\mathcal{S}}_1-7\bar{\mathcal{S}}_2}{3}&0\\
4\bar{\mathcal{T}}-\bar{\mathcal{S}}_2&0\\
\end{array}\!\!\right)\right\},
\end{align}
and
\begin{align}
    &\mathbf{K}{_{\TPV}}^{\frac{3}{2}}_{1\frac{3}{2}, 0 \frac{3}{2}}(k,p,E)=\frac{8\sqrt{10}\pi i} {3kp}\left(pQ_0(a)-kQ_1(a)\right)\left(\!\!\begin{array}{cc}
\frac{\bar{\mathcal{S}}_1-\bar{\mathcal{S}}_2}{3}&0\\
0&0
\end{array}\!\!\right).
\end{align}
Using the fact that our interactions are $T$-odd the time reversed version of these kernels is given by
\begin{equation}
    {\mathbf{K}_{\TPV}}^{J}_{L'S',LS}(k,p,E)=\left[{\mathbf{K}_{\TPV}}^{J}_{LS,L'S'}(p,k,E)\right]^\dagger,
\end{equation}
where the $\dagger$ takes the conjugate transpose of the c.c.~space matrix.

\section{\label{sec:observables}Observables}

The relation between the $N\!d$ scattering amplitude in the spin basis and partial wave basis is given by
\begin{align}
    \label{eq:basisrelation}
    M_{m_1'm_2',m_1m_2}=\sqrt{4\pi}\sum_{\beta}\sqrt{2L+1}\CG{L}{S}{J}{0}{m_S}{M}\CG{L'}{S'}{J}{m_L'}{m_S'}{M}\CG{1}{\frac{1}{2}}{S}{m_1}{m_2}{m_S}\CG{1}{\frac{1}{2}}{S'}{m_1'}{m_2'}{m_S'}{Y_{L'}^{m_L'}}^*(\hat{p})M^{J}_{L'S',LS},
\end{align}
where $m_1$ ($m_2$) is the initial spin of the deuteron (nucleon) and $m_1'$ ($m_2'$) is the final spin of the deuteron (nucleon).  The sum $\beta$ is over all indices other than $m_1$, $m_2$, $m_1'$, and $m_2'$.  At low energies it is sufficient to truncate the sum to values of $L=0$ or $1$ and $L'=0$ or $1$.  In the partial wave basis the $N\!d$ scattering amplitude is given by
\begin{equation}
    M^{J}_{L'S',LS}=Z_{\mathrm{LO}}t^{J;Nt\to Nt}_{L'S',LS}(k,k,E),
\end{equation}
where $t^{J;Nt\to Nt}_{L'S',LS}(k,k,E)$ is understood to be either $T\!P$ or $\TPV$. One set of $\TPV$ observables is given by the correlation $\vectS{\sigma}_N\cdot(\vect{k}\times\vectS{\epsilon}_d)$, where $\vectS{\sigma}_N$ is the spin of the nucleon, $\vect{k}$ is the momentum of the incoming nucleon beam, and $\vectS{\epsilon}_d$ is the polarization of the deuteron.  Choosing $\vect{k}$ to be along the $z$-axis and the deuteron polarization to be along the $y$-axis, the difference in cross sections for the nucleon polarized along and opposite the $\vect{k}\times\vectS{\epsilon}_d$ axis is given by\footnote{Note, although using the polarization conventions of Song et \emph{al.}~\cite{Song:2011sw} our expressions for the observables seem to differ from Song et \emph{al.} by an overall sign.  Their phase seems to coincide with using the density matrix method~\cite{glockle2012quantum,Fukukawa:2011wu}.}
\begin{align}
    &\Delta \sigma=\left(\frac{M_N}{3\pi}\right)^2\sum_{m_1',m_2'}\int\!\! d\Omega\,\frac{1}{8}\left\{\left|\sum_{m_1,m_2}\!\!f(m_1)(-1)^{\frac{1}{2}-m_2}M_{m_1'm_2',m_1,m_2}\right|^{\,2}\right.\\\nonumber
    &\hspace{7cm}\left.-\left|\sum_{m_1,m_2}\!\!f(m_1)M_{m_1'm_2',m_1m_2}\right|^{\,2}\right\},
\end{align}
while the sum of cross-sections is given by
\begin{align}
    &\sigma=\left(\frac{M_N}{3\pi}\right)^2\sum_{m_1',m_2'} \int\!\! d\Omega\,\frac{1}{8}\left\{\left|\sum_{m_1,m_2}\!\!f(m_1)(-1)^{\frac{1}{2}-m_2}M_{m_1'm_2',m_1,m_2}\right|^{\,2}\right.\\\nonumber
    &\hspace{7cm}\left.+\left|\sum_{m_1,m_2}\!\!f(m_1)M_{m_1'm_2',m_1m_2}\right|^{\,2}\right\},
\end{align}
where
\begin{equation}
    f(m_1)=\left\{\begin{array}{rl}
         1,& \mathrm{if\,} m_1=1  \\
         i\sqrt{2},& \mathrm{if\,}m_1=0  \\
         -1,& \mathrm{if\,}m_1=-1
    \end{array}\right..
\end{equation}
Using the expression for $M_{m_1'm_2',m_1m_2}$ in the partial wave basis the spin correlation coefficient $\Delta\sigma/(\sigma)$ is given by
\begin{align}
    A_{xy}=\label{eq:TVaymmetry}
    &\frac{3}{2}\mathrm{Im}\left[\sqrt{2}M^{\frac{1}{2}}_{0\frac{1}{2},0\frac{1}{2}}\left(M^{\frac{1}{2}}_{0\frac{1}{2},1\frac{3}{2}}\right)^*+\sqrt{2}M^{\frac{1}{2}}_{1\frac{3}{2},0\frac{1}{2}}\left(M^{\frac{1}{2}}_{1\frac{3}{2},1\frac{3}{2}}\right)^*\right.\\\nonumber
    &\hspace{2cm}\left.+2M^{\frac{3}{2}}_{0\frac{3}{2},0\frac{3}{2}}\left(M^{\frac{3}{2}}_{0\frac{3}{2},1\frac{1}{2}}\right)^*+2M^{\frac{3}{2}}_{1\frac{1}{2},0\frac{3}{2}}\left(M^{\frac{3}{2}}_{1\frac{1}{2},1\frac{1}{2}}\right)^*\,\right]/\\\nonumber
    &\hspace{2cm}\left(\left|M^{\frac{1}{2}}_{0\frac{1}{2},0\frac{1}{2}}\right|^2+2\left|M^{\frac{3}{2}}_{0\frac{3}{2},0\frac{3}{2}}\right|^2+3\left|M^{\frac{1}{2}}_{1\frac{1}{2},1\frac{1}{2}}\right|^2+6\left|M^{\frac{1}{2}}_{1\frac{3}{2},1\frac{3}{2}}\right|^2\right).
\end{align}
If $T$-symmetry is not violated then it can be seen that this asymmetry is zero as expected.  Below the deuteron breakup threshold this observable can also be calculated using the optical theorem which gives
\begin{align}
    &\Delta \sigma=\frac{4M_N}{3k}& \frac{1}{8}\mathrm{Im}\left\{\sum_{m_1,m_2}\sum_{m_1,m_1'}f(m_1)f^*(m_1')((-1)^{1-m_2-m_2'}-1)M_{m_1'm_2',m_1m_2}\Big|_{\theta=0}\right\},\\\nonumber
\end{align}
for $\Delta\sigma$ and 
\begin{align}
    &\sigma=\frac{4M_N}{3k}\frac{1}{8}\mathrm{Im}\left\{\sum_{m_1,m_2}\sum_{m_1,m_1'}f(m_1)f^*(m_1')((-1)^{1-m_2-m_2'}+1)M_{m_1'm_2',m_1m_2}\Big|_{\theta=0}\right\},\\\nonumber
\end{align}
for $\sigma$, where $\theta=0$ means only the forward scattering amplitude is taken.  Using these relationships from the optical theorem and plugging in the expression for the amplitudes in the partial wave basis gives
\begin{equation}
A_{xy}=\frac{3}{4}\frac{ \mathrm{Re}\left[ \sqrt{2} M^{\frac{1}{2}}_{0\frac{1}{2},1\frac{3}{2}}- \sqrt{2}
   M^{\frac{1}{2}}_{1\frac{3}{2},0\frac{1}{2}}+2
   M^{\frac{3}{2}}_{0\frac{3}{2},1\frac{1}{2}}-2
   M^{\frac{3}{2}}_{1\frac{1}{2},0\frac{3}{2}}\right]}{
   \mathrm{Im}\left[M^{\frac{1}{2}}_{0\frac{1}{2},0\frac{1}{2}}+2
   M^{\frac{3}{2}}_{0\frac{3}{2},0\frac{3}{2}}+3
   M^{\frac{1}{2}}_{1\frac{1}{2},1\frac{1}{2}}+6
   M^{\frac{1}{2}}_{1\frac{3}{2},1\frac{3}{2}}\right]}
\end{equation}
for the spin correlation coefficient $\Delta\sigma/\sigma$.  This expression is only valid for energies below the deuteron breakup threshold and at these energies is found to be equivalent to results from Eq.~(\ref{eq:TVaymmetry}), which serves as a check on our results.

Another possible $\TPV$ observable is the spin rotation of the neutron through a polarized deuteron target, with rotation angle $\phi$, about the axis $\vect{k}\times\vectS{\epsilon}_d$ given by the expression~\cite{Stodolsky:1981vn}
\begin{align}
    \frac{d\phi}{dz}=-\frac{2M_N N}{3k}\frac{1}{8}\sum_{m_1',m_2'}\sum_{m_1,m_2}& \mathrm{Re}\left\{f(m_1)f^*(m_1')((-1)^{\frac{1}{2}-m_2-m_2'}-1)M_{m_1',m_2',m_1,m_2}\Big|_{\theta=0}\right\}.
\end{align}
$N$ is the number of atoms per unit volume, $z$ is the length of the target through which the neutron travels, and $k$ is the momentum of the neutron in the c.m.~frame.  Plugging in Eq.~(\ref{eq:basisrelation}) the spin rotation per unit length in the partial wave basis is
\begin{equation}
    \frac{d\phi}{dz}=\frac{M_N N}{3k}\mathrm{Im}\left[ \sqrt{2}M^{\frac{1}{2}}_{0\frac{1}{2},1\frac{3}{2}}-
   \sqrt{2}M^{\frac{1}{2}}_{1\frac{3}{2},0\frac{1}{2}}+ 2
   M^{\frac{3}{2}}_{0\frac{3}{2},1\frac{1}{2}}- 2
   M^{\frac{3}{2}}_{1\frac{1}{2},0\frac{3}{2}}\right].
\end{equation}

\section{\label{sec:results}Results}

The spin correlation coefficient $A_{xy}$ for $N\!d$ scattering in the large-$N_C$ basis of LECs is
\begin{equation}
\label{eq:Axy}
A_{xy}=\tau_3\bar{g}_1^{(N_C)}A_{xy}^{(1)}+\bar{g}_2^{(N_C^0)}A_{xy}^{(2)}+\bar{g}_3^{(N_C^0)}A_{xy}^{(3)}+\tau_3\bar{g}_5^{(N_C^{-1})}A_{xy}^{(5)}.
\end{equation}
For $pd$ ($nd$) scattering $\tau_3=1$ ($\tau_3=-1$).  The values of $A_{xy}^{(i)}$ for each LEC at various nucleon lab energies are given in Table~\ref{tab:correlation}.
\begin{table}[hbt!]
    \centering
    \begin{tabular}{|c|ccccc|}\hline
    $E_{\mathrm{lab}}$ [MeV] & 0.225 & 1 & 2 & 3 & 5 \\\hline
    $A_{xy}^{(1)}$ [MeV] & 39.7 & 64.5 & 71.0 & 72.5 & 79.6 \\
    $A_{xy}^{(2)}$ [MeV] & \hspace{2mm}64.7\hspace{2mm} & 118\hspace{2mm} & 152\hspace{2mm} & 179\hspace{2mm} & 238\hspace{2mm} \\
    $A_{xy}^{(3)}$ [MeV] & -10.2 & -22.4 & -36.6 & -50.5 & -75.9 \\
    $A_{xy}^{(5)}$ [MeV] & 53.4 & 94.4 & 120 & 140 & 181 \\\hline
    \end{tabular}
    \caption{\label{tab:correlation}Coefficients $A_{xy}^{(i)}$ in front of each LEC for spin correlation coefficient (see Eq.~(\ref{eq:Axy})) for various nucleon lab energies.}
\end{table}
To factor out the large-$N_C$ dependence of each LEC we divide $A_{xy}^{(i)}$ by the appropriate large-$N_C$ scaling and normalize by the largest value of $|A_{xy}^{(i)}|$ scaled by large-$N_C$ to compare the respective contributions to $A_{xy}$ on an equal footing.\footnote{After rescaling by factors of $N_C$ $A_{xy}^{(1)}$ gives the largest contribution for the nucleon lab energies considered}  This procedure gives the results in Table~\ref{tab:normalized}.
\begin{table}[hbt!]
    \centering
    \begin{tabular}{|c|ccccc|}\hline
    $E_{\mathrm{lab}}$ [MeV] & 0.225 & 1 & 2 & 3 & 5 \\\hline
    $A_{xy}^{(1)}/|A_{xy}^{(1)}|$ & 1.00 & 1.00 & 1.00 & 1.00 & 1.00 \\
    $A_{xy}^{(2)}/(N_C|A_{xy}^{(1)}|)$  & 0.543 & 0.611 & 0.712 & 0.823 & 0.996 \\
    $A_{xy}^{(3)}/(N_C|A_{xy}^{(1)}|)$ & \hspace{2mm}-0.0860\hspace{2mm} & \hspace{2mm}-0.116\hspace{2mm} & -0.172 & -0.232 & -0.318 \\
    $A_{xy}^{(5)}/(N_C^2|A_{xy}^{(1)}|)$ & 0.149 & 0.163 & \hspace{2mm}0.187\hspace{2mm} & \hspace{2mm}0.214\hspace{2mm} & \hspace{2mm}0.252\hspace{2mm} \\\hline
    \end{tabular}
    \caption{\label{tab:normalized}Coefficients $A_{xy}^{(i)}$ for various nucleon lab energies normalized by factors of $N_C$ ($N_C=3$) and by $|A_{xy}^{(1)}|$.}
\end{table}
It is apparent that the contribution from the LO($\mathcal{O}(N_C)$) in large-$N_C$ LEC $\bar{g}_1^{(N_C)}$ dominates $A_{xy}$.  The NLO($\mathcal{O}(N_C^0)$) $g_2^{(N_C^0)}$ also gives a significant contribution, about half as much as $\bar{g}_1^{(N_C)}$ at lower energies and at higher energies is comparable to the contribution from $\bar{g}_1^{(N_C)}$. Meanwhile the NNLO($\mathcal{O}(N_C^{-1})$) in large-$N_C$ LEC $\bar{g}_5^{(N_C^{-1})}$ and the NLO($\mathcal{O}(N_C^{-1})$) in large-$N_C$ LEC $\bar{g}_3^{(N_C^{0})}$ each contribute $\sim$20\% of the leading contribution from $\bar{g}_1^{(N_C)}$ at higher energies.  Thus, using large-$N_C$ counting we find that $A_{xy}$ to LO in \EFT is predominantly determined by $g_1^{(N_C)}$ and $g_2^{(N_C^0)}$.

The spin rotation of the neutron through a polarized deuteron target gives the prediction
\begin{equation}
    \frac{1}{N}\frac{d\phi}{dz}=\left(-2.22\,\frac{\mathrm{rad}}{\mathrm{MeV}}\right)\left[0.407\bar{g}_1^{(N_C)}-0.804\bar{g}_2^{(N_C^0)}+0.445\bar{g}_3^{(N_C^0)}+\bar{g}_5^{(N_C^{-1})}\right]
\end{equation}
To obtain this value we calculated the spin rotation observables for small c.m.~momentum approaching zero momentum until it was found to converge.  This value comes from a c.m.~momentum of $k_{\mathrm{cm}}=0.1$ keV.  Normalizing the spin rotation such that all LECs have the same large-$N_C$ scaling gives
\begin{equation}
    \frac{1}{N}\frac{d\phi}{dz}=\left(-0.904\,\frac{\mathrm{rad}}{\mathrm{MeV}}\right)\left[\bar{g}_1^{(N_C)}-0.659(N_C \bar{g}_2^{(N_C^0)})+0.364(N_C \bar{g}_3^{(N_C^0)})+0.273(N_C^2 \bar{g}_5^{(N_C^{-1})})\right],
\end{equation}
where $N_C=3$ is used.  It is apparent that the LO($\mathcal{O}(N_C)$) in large-$N_C$ LEC $\bar{g}_1^{(N_C))}$ gives the largest contribution to the spin rotation, while the NLO($\mathcal{O}(N_C^{0})$) in large-$N_C$ LEC $\bar{g}_2^{(N_C^0)}$ gives a smaller but significant contribution of $\sim$70\% the leading contribution.  The NNLO($\mathcal{O}(N_C^{-1})$) in large-$N_C$ LEC $\bar{g}_5^{(N_C^{-1})}$ and the NLO($\mathcal{O}(N_C^0)$) in large-$N_C$ LEC $\bar{g}_3^{(N_C^0)}$ each contribute $\sim$30\% of the leading contribution.  Note, the spin rotation is sensitive to the same LECs as the spin correlation coefficient.  None of the observables depend on the LEC $\bar{g}_4^{(N_C^0)}$ since it corresponds to a $\Delta I=2$ operator which cannot connect an isospin-$\frac{1}{2}$ state to itself without violating isospin, which occurs beyond LO in \EFT and its contribution is thus suppressed.

\section{\label{sec:conclusion}Conclusion}

At low energies $\TPV$ interactions can be described in terms of $\NN$ contact interactions by five LECs.  Building on the large-$N_C$ analyses of Refs.~\cite{Samart:2016ufg,Schindler:2015nga}, we showed that a linear combination of the isovector LECs is $\mathcal{O}(N_C)$, the two isoscalar LECs and one isotensor LEC are $\mathcal{O}(N_C^0)$, and a linear combination of the isovector LECs is $\mathcal{O}(N_C^{-1})$.  Isoscalar LECs receive contributions from both the $\bar{\theta}$ term of QCD and $d=6$ $\cpv$ SM EFT operators, while the isovector and isotenstor terms only receive contributions from $d=6$ $\cpv$ SM EFT operators~\cite{Maekawa:2011vs}. Thus the pattern predicted by large-$N_C$ considerations could be broken if the contributions from the $\bar{\theta}$ term and $d=6$ $\cpv$ SM EFT operators are sufficiently disparate.  This is also means that measuring each invidual LEC can help to disentangle contributions from the $\bar{\theta}$ term and $d=6$ $\cpv$ SM EFT operators to $CP$ violating nuclear observables.

We did not consider $\Tv P$-conserving interactions in this work as they contain an extra power of momentum and are thus even more suppressed as compared to $\TPV$ intearctions at low energies.  However, plans are underway for an experiment  using $pd$ scattering at the cooler synchrotron COSY facility, with both polarized beam and target, to investigate  $\Tv P$ interactions~\cite{Lenisa:2019cgb,Eversheim:2017zxl,Valdau:2016ols}.  Results from this experiment would warrant future investigations of $\Tv P$ interactions in \EFT.

$\TPV$ $\NN$ interactions in the $N\!d$ system at low energies are sufficiently described by three-nucleon $S$- to $P$-wave transition amplitudes.  Calculating all such transition amplitudes we investigated the $\TPV$ observables  of neutron spin rotation through a polarized deuteron target and a spin correlation coefficient in $N\!d$ scattering.  Both observables are related to the correlation $\vectS{\sigma}_N\cdot(\vect{k}\times\vectS{\epsilon}_d)$.  At LO in \EFT these observables depend on four of the five LECs since $\bar{g}_4^{(N_C^0)}$ requires isospin violation, which occurs at higher orders in \EFT.  Putting these observables in the large-$N_C$ basis we find that both the spin-rotation and spin correlation coefficient are predominantly determined by the LO($\mathcal{O}(N_C)$) in large-$N_C$ LEC $\bar{g}_1^{(N_C)}$ and NLO($\mathcal{O}(N_C^0)$) in large-$N_C$ LEC $\bar{g}_2^{(N_C^0)}$, with the spin rotation being more sensitive to $\bar{g}_1^{(N_C)}$.  At the neutron lab energies considered in this work contributions from $P$ to $D$-wave transition amplitudes should not be significant as was found in the PV case~\cite{Vanasse:2018buq}.  In addition our calculations did not consider Coulomb interactions.  However, at higher energies Coulomb interactions give perturbative corrections of the size $\alpha M_N/p$.  At nucleon lab energies of $E_{\mathrm{lab}}=1$ MeV Coulomb corrections give a $\sim\!\!24\%$ correction, while at $E_{\mathrm{lab}}=5$ MeV they give a $\sim\!\!11\%$ correction.  The latter correction is roughly on par with the size of NLO corrections in \EFT. A NLO calculation of these $\TPV$ violating observables will likely require the inclusion of a $\TPV$ violating three-body force.  As shown by Vanasse~\cite{Vanasse:2018buq} in contradiction to the work of Grie{\ss}hammer and Schindler~\cite{Griesshammer:2010nd} a NLO PV three-body force is required by RG arguments.  The similarity of the PV $\NN$ interactions to the $\TPV$ $\NN$ interactions suggests that the necessity for a NLO PV three-body force implies the need for a NLO $\TPV$ three-body force.

\EFT has also been used to investigate the deuteron, triton, and $^3$He EDMs, associated radii, and form factors~\cite{Yang:2020ges}.  These calculations required the nucleon EDMs which cannot be directly calculated in \EFT, but must be included as input either from experiment or $\chi$EFT~\cite{Maekawa:2011vs}.  Matching \EFT to $\chi$EFT, predictions for light nuclei can be made using the simpler formalism of \EFT in terms of $\chi$EFT parameters.  This also avoids the complication of RG non-invariance in $\chi$EFT~\cite{Valderrama:2016koj}, which does not exist in \EFT.  Few nucleon EDMs and the observables of this work offer a suite of possible nuclear osbservables to measure the five LECs that describe low energy $\cpv$
nuclear interactions and can be used to understand contributions from BSM physics. 

\acknowledgments{We would like to thank Hersh Singh for useful suggestions on notation.  We also want to thank E. Mereghetti, M.R. Schindler, and W.M. Snow for useful discussions and comments that led to improvements and corrections to the manuscript.}

\appendix

\end{document}